\documentclass[twocolumn]{article}
\usepackage{icqproc,epsf,amssymb,amsmath}

\newcommand{\bs}[1]{\boldsymbol{#1}}

\begin{document}
\thispagestyle{empty}

\twocolumn[
\vspace*{30mm}
\begin{LARGE} 
\begin{center}
%
%
%
Diffraction of the Dart-Rhombus Random Tiling
%
\end{center}
\end{LARGE}
\begin{large}
\begin{center} 
%
%
Moritz H\"offe
%
%
\end{center}
\end{large}
\begin{footnotesize}
\begin{it}
\begin{center}
%
%
Institut f\"ur Theoretische Physik, Universit\"at T\"ubingen, 
Auf der Morgenstelle 14, D-72076 T\"ubingen, Germany  
%
%
\end{center}
\end{it}
\end{footnotesize}
\begin{footnotesize}
\begin{center}
%
%
%
%
\end{center}
\end{footnotesize}
\vspace{4ex}
\begin{small}
\hrule\vspace{3ex}
\begin{minipage}{\textwidth}
{\bf Abstract}\vspace{2ex}\\
\hp 
%
%
The diffraction spectrum of the dart-rhombus random tiling of the
plane is derived in rigorous terms. Using the theory of dimer models, 
it is shown that it consists of Bragg peaks and an
absolutely continuous diffuse background, but no singular continuous 
component. The Bragg part is given explicitly.
%
%
\vspace{2.5ex}
\end{minipage}\vspace{3ex}
\hrule
\end{small}\vspace{6ex}
]

%
%

\section{Introduction}

Structure models of quasicrystals are usually based on the
asumption of either energetical or entropical stabilization of
the material.  Random tilings as possible structure models 
of the second kind where proposed \cite{Elser} soon after 
the discovery of quasicrystals and studied thoroughly since,
see e.g.\ \cite{Henley91,Richard,Gier} and references therein.
Nonetheless, until today it is not yet clear which mechanism is
dominating, although experiments are indicating a stochastic
component in many cases \cite{JRB}.  Scaling arguments
\cite{Henley88,Henley91} predict a singular continuous contribution to 
the diffraction spectrum for two-dimensional random tilings in addition 
to the usual Bragg part and continuous background. This should
be visible in diffraction images of materials with 
so-called T-phases, see \cite{Baake} and references therein,
though it is not obvious how to distinguish the different
contributions. This underlines the necessity of 
investigating the diffraction of random tilings in more
detail.   

In this article, we illustrate recently established results
\cite{BH} by the so-called dart-rhombus tiling. This
two-dimensional model has crystallographic symmetries and can be
mapped onto the dimer model on the Fisher lattice. After
introducing the tiling and the necessary mathematical tools, we 
calculate the two-point correlation
functions and thereof the diffraction spectrum. As in other 
crystallographic examples, the spectrum 
can be shown to consist only of a Bragg part and an absolutely
continuous background, i.e.\ there is no singular continuous component.

\section{The dart-rhombus random tiling}

\begin{figure}[ht]
\centerline{\epsfxsize=6cm \epsfbox{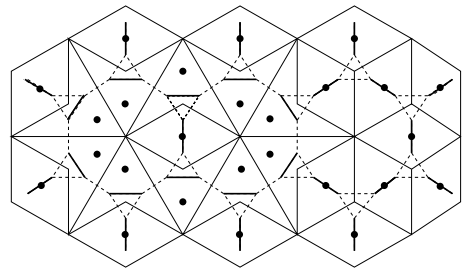}}
\caption{The dart-rhombus tiling as 
dimer model on the Fisher lattice. The dots represent the atomic 
scatterers.}
\label{tiling}
\end{figure}

\begin{figure}[ht]
\centerline{
\begin{minipage}[t]{3cm} 
\vspace{-3cm}\epsfxsize=3cm \epsfbox{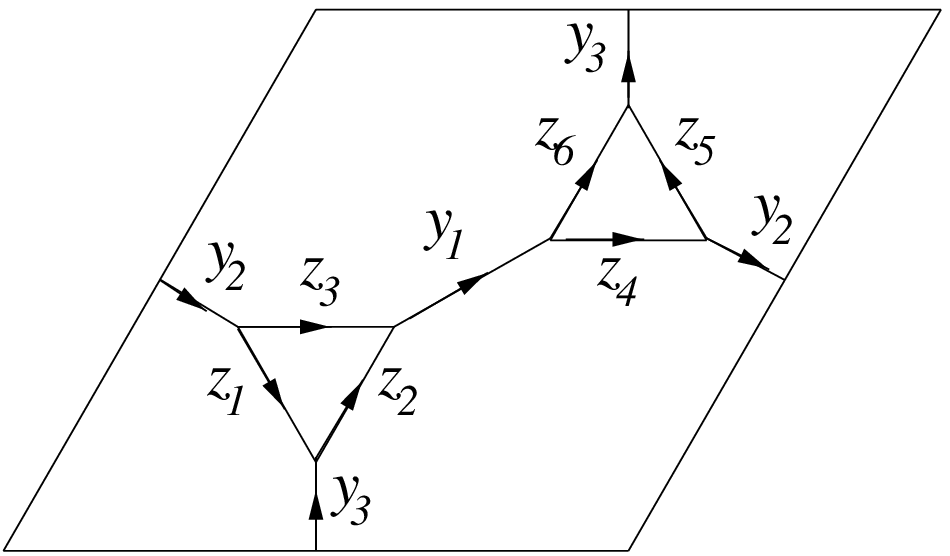}
\end{minipage}
\epsfxsize=5cm \epsfbox{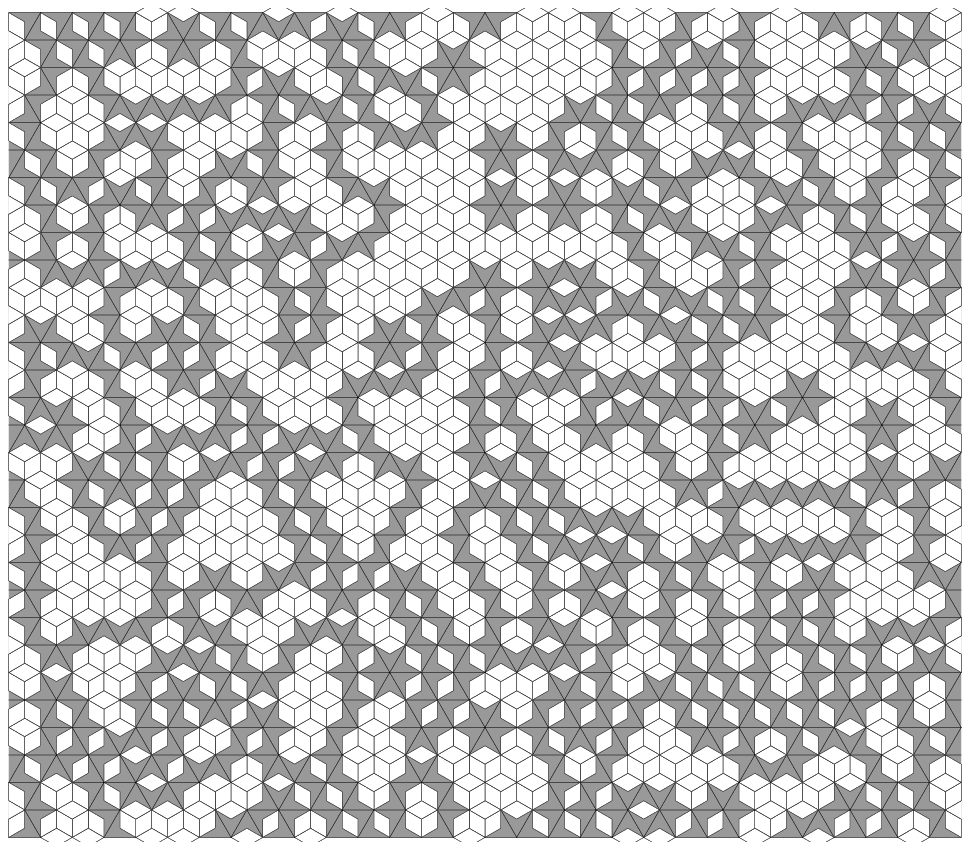}}
\caption{Elementary cell of the Fisher lattice with activities 
assigned to each bond and a typical random tiling with 
$\rho_1=0.21$, $\rho_2=0.19$, $\rho_3=0.17$.} \label{elemFisher}
\end{figure}

The dart-rhombus tiling is a filling of the plane, without
gaps and overlaps, with $60^{\circ}$-rhombi of side $1$ and darts
made of two rhombus halves (Fig.\ \ref{tiling}).
In addition to the usual face-to-face condition, we impose an 
alternation condition on the rhombi, such that neighbouring 
rhombi of equal orientation are excluded. Finally, to avoid 
pathological lines of alternating darts, we demand that two 
neighbouring darts must not share a short edge. These rules 
force the darts to form closed loops in a background of 
alternating rhombi. The minimal total rhombus density  
obviously is $1/3$. This tiling can be mapped onto the fully packed 
dimer model on an Archimedian tiling known as Fisher's lattice 
in the context of statistical mechanics. In order to control 
the densities of the different prototiles, we weigh them using 
activities $y_i,z_i$ (Fig.\ \ref{elemFisher}), with 
$z_i=e^{\beta\mu_i^{}}$ etc., 
where $\mu_i^{}$ are chemical potentials and $\beta$ the inverse 
temperature.

The grand-canonical partition function is therefore equal to the 
dimer generating function. For any periodic graph with even number 
of sites in the elementary cell, the latter can be computed as 
Pfaffian\footnote{This is basically the square root of the 
determinant of an even 
antisymmetric matrix \cite[Ch.\ IV.2]{MW}.} of the suitably 
activity-weighted adjacency matrix $\bs A$ \cite{Kasteleyn}. 
The calculation of the 
Pfaffian is simplified considerably by imposing periodic 
boundary conditions, but in the infinite volume limit the 
result holds also for free ones \cite[Lemma 1]{BH}.

Let us denote the rhombus densities by $\rho_i^{}$ ($i=1,2,3$) 
and the the dart densities by $\sigma_j^{}$ ($j=1,\dots,6$). 
There are several constraints on the densities. Closed dart 
loops require equal densities of opposite darts:
\begin{equation} \label{constsigma}
\sigma_1^{}=\sigma_5^{}, \quad \sigma_2^{}=\sigma_6^{}, 
\quad \sigma_3^{}=\sigma_4^{}.
\end{equation}
Moreover, as each dart is accompanied by a corresponding rhombus, 
the remaining rhombi occur with equal frequency owing to the 
alternation condition,
\begin{equation} \label{constrho}
\rho_1^{}-\sigma_1^{}=\rho_2^{}-\sigma_2^{}=\rho_3^{}-\sigma_3^{}.
\end{equation}
Including the normalization constraint (the sum of the densities 
is $1$), the number of independent parameters (activities or 
densities) reduces to three. We exploit this freedom by setting 
all activities except $z_1,z_2$ and $z_3$ equal to $1$. The 
dart-rhombus tiling undergoes second order phase transitions at
\begin{equation}
\begin{align}
1+z_1^2+z_2^2+z_3^2&=2\max\{1,z_1^2,z_2^2,z_3^2\}\;\text{or} \nonumber\\
z_1^2+z_2^2+z_3^2&=2 \max \{z_1^2,z_2^2,z_3^2\}
\end{align}
\end{equation}
with logarithmic (Onsager type) or square root divergence (Kasteleyn
type), respectively. The point of maximum entropy is fixed by symmetry to 
$\rho_i^{}=\frac{1}{6}$, $\sigma_j^{}=\frac{1}{12}$, where darts and rhombi 
occupy half of the tiling area each. For further 
details see \cite{Richard,Hoeffe}. 

\section{Diffraction theory}

For simplicity, we assume kinematic diffraction in the Fraunhofer 
picture \cite{Cowley}, i.e.\ diffraction at infinity from 
single-scattering. 
The diffracted intensity
$\widehat{\gamma}_{\omega}$ (a positive measure) is calculated as 
Fourier transform of the autocorrelation $\gamma_{\omega}$ (see \cite{BH} for details).
It is known that every positive measure
admits a unique decomposition into three parts $\mu=\mu_{pp}^{}+\mu_{sc}^{}
+\mu_{ac}^{}$ with respect to Lebesgue's measure, where $pp$, $sc$ and $ac$ 
stand for pure point, singular continuous and absolutely continuous 
\cite{RS}. In a diffraction spectrum, $\mu_{pp}^{}$ are the Bragg peaks 
and $\mu_{ac}^{}$ the usual diffuse background or Laue scattering. A singular 
continuous part can be encountered in 1D substitutional sequences,  
cf \cite{Enter}, and is also expected for 2D quasicrystalline random tilings
\cite{Henley88,Henley91,BH}.

Consider the so-called weighted Dirac comb \cite{Cordoba}
\begin{equation}
\omega=\sum_{x\in\tilde{\Gamma}}
w(x)\delta_x
\end{equation}
on a lattice $\tilde{\Gamma}$, where $\delta_x$ is the unit point
measure (Dirac measure) concentrated at $x$, and $w(x)\in\{0,1\}$ is chosen 
in order to obtain
a specific member of the random tiling ensemble. 
Its autocorrelation $\gamma_{\omega}$ is (almost surely)
\begin{equation} \label{auto}
\gamma_{\omega}=\sum_{z\in\Delta} \nu(z) \delta_z.
\end{equation}
Here, $\Delta=\tilde{\Gamma}-\tilde{\Gamma}=\tilde{\Gamma}$ is the set of 
difference vectors and
the autocorrelation coefficient $\nu(z)$ can be calculated 
according to 
\begin{equation}
   \nu(z) \; = \; \lim_{R\to\infty} \frac{1}{{\rm vol}(B_R)}
          \sum_{\stackrel{\scriptstyle y \in \Lambda_R}
                         {\scriptstyle y+z \in \Lambda}}
               \overline{w(y)} \, w(y+z) \, ,
\end{equation}
where $B_R$ is the ball of radius $R$ around the origin,
$\Lambda_R=\Lambda\cap B_R$ and \raisebox{1ex}{$\overline{\hphantom{x}}$}
the complex conjugation.
Thus, $\nu(z)$ is simply the probability of having two scatterers
at distance $z$, which a.s.\ exists.

We use standard Fourier theory of tempered distributions, 
for our conventions see \cite{BH}.

\section{Diffraction of the dart-rhombus tiling}

We decorate the tiling with point scatterers 
 $\delta_x$ according to
Fig.\ \ref{tiling}. More realistic atomic profiles can be handled 
with the convolution theorem \cite[Ch.\ IX]{RS}. The scatterers may have 
complex strengths $h_{\rho_i}$, resp.\ $h_{\sigma_j}$, where the 
strengths of opposite darts are supposed to be equal. This constraint
simplifies the calculations but is not necessary.
The point set of
all possible atomic positions is a Kagom\'e grid with minimal vertex 
distance $1/2$. We write it as triangular lattice $\Gamma$ with a
rhombic elementary cell $E$ containing the nine scatterers positions 
for the different tiles (cf Fig.\ \ref{elemFisher}). Introducing basis 
vectors $e_1=(\sqrt{3},0)^t$, 
$e_2=1/2(\sqrt{3},3)^t$ for $\Gamma$ and $E$, the positions $p$ in 
$E$
(with corresponding density) are given by ($a=1/4(\sqrt{3},1)^t$, 
$b=1/4(\sqrt{3},-1)^t$)
\begin{align} \label{pos}
  p_{\rho_1^{}}=&\,3a,& p_{\rho_2^{}}=&\,2a-b,& 
  p_{\rho_3^{}}=&\,a+b, \nonumber\\
  p_{\sigma_1^{}}=&\,a,& p_{\sigma_2^{}}=&\,2a+b,& 
  p_{\sigma_3^{}}=&\,3a-b, \\
  \quad p_{\sigma_4^{}}=&\,3a+b,& p_{\sigma_5^{}}=&\,5a,&
   p_{\sigma_6^{}}=&\,4a-b. \nonumber
\end{align}

We now have to
calculate the autocorrelation or the joint
occupation probability of the dimers. As we will see,
the autocorrelation coefficients
can be split into a constant term and one decreasing with the distance
between the scatterers. After taking the Fourier transform, the 
first will yield
the Bragg peaks whereas the second will be responsible for the continuous
part of the spectrum; we will show that this can be represented
by a continuous function and hence contains no singular contribution.

Using Gibbs' weak phase rule \cite{Ruelle,BH}, one can prove the ergodicity of the
model and thus justify the calculation of the diffraction spectrum 
via the ensemble average. 

Let $\eta^{}_{\bs{kk}'}$ be the occupation variable that takes the value
$1$ if the bond between $\bs{k}$ and $\bs{k}'$ is occupied and $0$ otherwise; 
$\bar{\eta}_{\bs{kk}'}^{}:=1-\eta^{}_{\bs{kk}'}$. As was shown in \cite{BH},
the probability $P_{\alpha\beta}$ of bonds $\alpha$ and 
$\beta$ being occupied simultaneously is given by
\begin{align} \label{auto2d}
    P_{\alpha\beta}  =&  \langle \eta^{}_{\bs{k}_{\alpha}^{}\bs{k}'_{\alpha}}
            \eta^{}_{\bs{k}_{\beta}^{}\bs{k}'_{\beta}}\rangle \nonumber\\      
        =&  \langle\eta^{}_{\bs{k}_{\alpha}^{}\bs{k}'_{\alpha}}\rangle\langle
         \eta^{}_{\bs{k}_{\beta}^{}\bs{k}'_{\beta}} \rangle  \nonumber\\
         &+\langle \bar{\eta}_{\bs{k}_{\alpha}^{}\bs{k}'_{\alpha}}
         \bar{\eta}_{\bs{k}_{\beta}^{}\bs{k}'_{\beta}}\rangle-\langle
         \bar{\eta}_{\bs{k}_{\alpha}^{}\bs{k}'_{\alpha}}\rangle\langle
         \bar{\eta}_{\bs{k}_{\beta}^{}\bs{k}'_{\beta}} \rangle \\
       =& \, \rho_{\alpha}^{}\, \rho_{\beta}^{}  \nonumber\\
       &-A_{\bs{k}_{\alpha}^{}\bs{k}'_{\alpha}}A_{\bs{k}_{\beta}^{}
       \bs{k}'_{\beta}}
        (A^{-1}_{\bs{k}_{\alpha}^{}\bs{k}_{\beta}^{}} A^{-1}_{\bs{k}'_{\alpha}
        \bs{k}'_{\beta}} -A^{-1}_{\bs{k}_{\alpha}^{}\bs{k}'_{\beta}}
        A^{-1}_{\bs{k}'_{\alpha}\bs{k}_{\beta}^{}}), \nonumber
\end{align}
with  
$\rho_{\alpha}^{}$ the 
density of dimers that can occupy the bond $(\bs{k}_{\alpha}^{}
\bs{k}'_{\alpha})$ and $\bs A$ the weighted adjacency matrix.

We consider the constant part of the autocorrelation first. Combining
(\ref{auto}) and (\ref{auto2d}) we get for the point set of scatterers with
density $2/\sqrt{3}$ 
\begin{equation}
(\gamma_{\omega}^{})_{const}^{}=
\omega_{\Gamma}^{}*\bigg(\frac{2}{\sqrt{3}}\sum_{\tau,\tilde{\tau}\in
\{\rho_i^{},\sigma_j^{}\}}\!\!\! (h_{\tau} h_{\tilde{\tau}}
\tau\tilde{\tau})\delta_{p_{\tau}-p_{\tilde{\tau}}}\bigg), \nonumber
\end{equation} 
where $*$ denotes convolution.
Computing the Fourier transform with Poisson's summation formula
\cite[Eq.\ 12]{BH} 
using (\ref{pos}) and (\ref{constsigma}), 
the pure point part of the spectrum is (almost surely)
\begin{align}
  (\widehat{\gamma})_{pp}=&\frac{4}{3}\sum_{(k,l)\in\Gamma^*}
  \bigg|h_{\rho_1}
  \rho_1^{}+(-1)^k h_{\rho_2}\rho_2^{}+(-1)^{l} 
  h_{\rho_3}\rho_3^{}  \nonumber\\
  &+2\cos{\textstyle \frac{\pi(k+l)}{3}}\Bigl((-1)^{k+l} 
  h_{\sigma_1}\sigma_1^{} \\
  &+(-1)^{l} h_{\sigma_2}\sigma_2^{}+(-1)^k h_{\sigma_3}\sigma_3^{}
   \Bigr)\bigg|^2 \delta_{(k,l)}^{}, \nonumber 
\end{align}
where $\Gamma^*$ is spanned by $e_1^*=\big(\frac{1}{\sqrt{3}},-\frac{1}{3}
\big)$, $e_2^*=\big(0,\frac{2}{3}\big)$.

It remains to calculate the other part of (\ref{auto2d}).
$A_{\bs{k}_{\alpha}^{} \bs{k}'_{\alpha}}$ is nonvanishing
only if $\bs{k}_{\alpha}^{}$ and $\bs{k}'_{\alpha}$ are connected; in 
this case $A_{\bs{k}_{\alpha}^{}\bs{k}'_{\alpha}}=\epsilon z_{\alpha}$, with 
$\epsilon=\pm 1$ according to the direction of the arrows in Fig.\ 
\ref{elemFisher}. 

Since $\bs{A}$ is the adjacency 
matrix of a graph that is an $(m,n)$-periodic array of elementary cells with 
toroidal boundary conditions and therefore cyclic, 
it can be reduced to the diagonal form $\bs{\Lambda}=\text{diag}\{\lambda_
{\bs j}\}$  by a Fourier-type similarity transformation with matrix  
$S_{\bs{kk}'}=
(mn)^{-1/2}\exp(2\pi i (k_1^{} k'_1/m +k_2^{} k'_2/n))$. $\bs{A}^{-1}$ is then 
determined by \cite{FS}
\begin{equation} \label{diag}
   A^{-1}_{\bs{kk}'}  =  
       \left(\bs{S\Lambda}^{-1}\bs{S}^{-1}\right)_{\bs{kk}'}
        =  \sum_{\bs{j}=(1,1)}^{(m,n)} S_{\bs{kj}}^{}
               \lambda_{\bs{j}}^{-1}S^{\dagger}_{\bs{k}'\bs{j}} \, .
\end{equation}
In the infinite volume limit, the sums approach integrals (Weyl's Lemma), 
and by introducing ${\bs r}=\bs{k}'-\bs{k}$, $\varphi_1^{}=2\pi i 
j_1^{}/m$ etc. we obtain
\begin{equation} \label{inverse}
     A^{-1}_{\bs{kk}'}  = 
         \frac{1}{4 \pi^2}\int\limits_0^{2 \pi}
         \int\limits_0^{2 \pi}\lambda^{-1}(\varphi_1^{},\varphi_2^{})
         e^{i\bs{\varphi}\cdot{\bs r}}
         d\varphi_1^{} d\varphi_2^{}.
\end{equation}
To determine $\lambda^{-1}$, observe that the inverse of 
\begin{equation}
\lambda=
\begin{pmatrix}
  0 & z_1 & z_3 & 0 & -e^{-i\varphi_1^{}} & 0 \\
  -z_1 & 0 & z_2 & 0 & 0 & -e^{-i\varphi_2^{}} \\
  -z_3 & -z_2 & 0 & 1 & 0 & 0 \\
  0 & 0 & -1 & 0 & z_3 & z_2 \\
  e^{i\varphi_1^{}} & 0 & 0 & -z_3 & 0 & z_1 \\
  0 & e^{i\varphi_2^{}} & 0 & -z_2 & -z_1 & 0
\end{pmatrix} \nonumber
\end{equation} 
can be computed easily by any computer-algebra package but unfortunately 
does not fit onto this page. 

Defining the coupling function $[x,y]_{p_1 p_2}$ for two dimers in elementary 
cells at distance $\bs{r} = x e_1 + y e_2$, with $(x,y)\in
\mathbb Z^2$, where the dimers occupy 
positions $p_1$, resp.\ $p_2$ in each elementary cell, we rewrite 
(\ref{inverse}) in more explicit form
\begin{equation} 
  [x,y]_{p_1 p_2}=\frac{1}{4 \pi^2}\int\limits_0^{2\pi}\int\limits_0^{2\pi}
  \frac{g(p_1,p_2,
  \varphi_1^{},\varphi_2^{})e^{i\bs \varphi\cdot\bs r}}{\det\left(
  \lambda(\varphi_1^{},\varphi_2^{})\right)}d\varphi_1^{} d\varphi_2^{},
  \nonumber
\end{equation}
where the determinant of $\lambda$ is given by
\begin{equation}
  \det(\lambda)=a+2b \cos\varphi_1^{}
  +2c\cos\varphi_2^{}+2d\cos(\varphi_1^{}-\varphi_2^{}), \nonumber
\end{equation}
with $a=z_1^4+z_2^4+z_3^4+1$, $b=z_1^2 z_2^2
-z_3^2$, $c=z_1^2 z_3^2-z_2^2$ and $d=z_2^2 z_3^2-z_1^2$; 
 $g$ can be taken from the corresponding entry in $\lambda^{-1}$ and
is finite. In order to determine the spectral type of the diffraction, we are 
interested in the asymptotic behaviour of $[x,y]_{p_1 p_2}$ for large 
$\bs r$ . 
Substituting $v=e^{-i\varphi_1^{}}$ and $w=e^{-i\varphi_2^{}}$, we obtain
that $I=4 \pi^2 [x,y]_{p_1 p_2}$ is
\begin{equation}
  \int\limits_{S^1\times S^1}
  \frac{-g(p_1,p_2,v,w) v^{-x}w^{-y}\;dv dw}
  {v^2(bw+d)+v(aw+c(w^2+1))+w(b+dw)},\nonumber
\end{equation}
with $S^1$ the unit circle.
\begin{figure}[ht]
\centerline{\epsfxsize=6cm \epsfbox{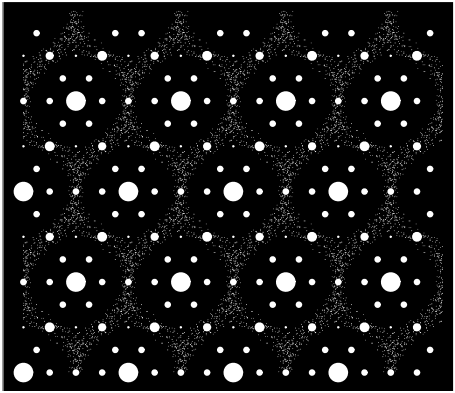}}
\caption{Diffraction image of the tiling of Fig.\ \ref{elemFisher}
with scatterers of equal strength $1$. It was calculated numerically by
means of standard FFT, because this is simpler than using the exact
expression for the $ac$ part.}
\label{diffr}
\end{figure}
We integrate over $v$ for $x < 0$; the case of positive $x$ 
can be treated analogously. The integrand is singular at
\begin{equation}
v_{\pm}=\frac{-\alpha\pm\sqrt{\alpha^2-4\beta}}
   {2(bw+d)}
\end{equation}
with $\alpha=(aw+c(w^2+1))$ and $\beta=w(bw+d)(b+dw)$, but only $v_+$ 
lies inside the unit circle. Thus,
\begin{equation} \label{int}
I = 2\pi i\int\limits_{S^1}\frac{g(p_1,p_2,v_+,w) v_+^{-x}w^{-y}}
  {(v_+-v_-)(bw+d)}dw.
\end{equation}
Away from the phase transitions, it can be shown that $|v_+|<1$. With 
$\tilde{v}_+=\max_{\varphi_2} v_+$ we get
\begin{align}
|I|&\leq 2\pi\int_{S^1} \frac{|g(p_1,p_2,v_+,w)||v_+|^{-x}}
  {|(v_+-v_-)(bw+d)|}dw \nonumber \\
  &\leq 2 \pi |\tilde{v}_+|^{-x} \int_{S^1}\frac{|g(p_1,p_2,v_+,w)|}
  {|(v_+-v_-)(bw+d)|}dw \nonumber \\
  &=\mathcal O\left(e^{-t_1 |x|}\right),   
\end{align}
for some positive constant $t_1$, because the remaining integral stays finite. 
Since the coupling function is invariant under interchange 
of $x$ and $y$, this can be shown for $y$ as well. As $I$ is maximal for $x=0$ 
for arbitrary but fixed $y$ and vice versa, we conclude that
\begin{equation}
[x,y]_{p_1 p_2}=\mathcal O\left(e^{-(t_1|x|+t_2|y|)}\right).
\end{equation}
The non-constant part of the correlation function in (\ref{auto2d}) 
consists basically of products of $[x,y]$. With such an asymptotic behaviour,
one can show that its Fourier transform indeed converges towards a 
continuous function on $\mathbb R^2/\Gamma$  (cf \cite[Addendum]{BH}). 

At the Kasteleyn 
phase transitions, we get crystals consisting only of one rhombus 
orientation and the corresponding darts (cf \cite{Richard}). The spectrum 
thus displays Bragg peaks only. For the Onsager case, we re-substitute
$w=e^{-i\varphi_2}$ in (\ref{int}). Because of the symmetry of the kernel, it is
sufficient to integrate from $0$ to $\pi$. The kernel reaches its maximum value 1 
only at $\varphi_2=0$ or $\pi$ and remains smaller elsewhere. Using the 
same argument as in the
treatment of the lozenge tiling in \cite{BH}, we estimate the kernel by a
decreasing/increasing straight line. From the resulting asymptotic behaviour we 
conclude that the diffuse part of the spectrum
is an absolutely continuous measure as well.

\section{Acknowledgement}

I am grateful to Michael Baake for valuable discussions and to 
Robert V.\ Moody and the Dept.\ of Mathematics, University of Alberta,
for hospitality, where part of this work was done.

%
%

\begin{footnotesize}
\begin{frenchspacing}

\end{frenchspacing}
\end{footnotesize}

\end{document}